\documentclass[twocolumn,english,aps,reprint]{revtex4}
\usepackage[T1]{fontenc}
\usepackage[latin9]{inputenc}
\usepackage{array}
\usepackage{amstext}
\usepackage{graphicx}

\makeatletter

\providecommand{\tabularnewline}{\\}

\@ifundefined{textcolor}{}
{%
 \definecolor{BLACK}{gray}{0}
 \definecolor{WHITE}{gray}{1}
 \definecolor{RED}{rgb}{1,0,0}
 \definecolor{GREEN}{rgb}{0,1,0}
 \definecolor{BLUE}{rgb}{0,0,1}
 \definecolor{CYAN}{cmyk}{1,0,0,0}
 \definecolor{MAGENTA}{cmyk}{0,1,0,0}
 \definecolor{YELLOW}{cmyk}{0,0,1,0}
}

\makeatother

\usepackage{babel}
\begin{document}
\title{Evidence of Slater-type mechanism as origin of insulating state in
Sr$_{2}$IrO$_{4}$}
\author{Vijeta Singh$^{1,2}$ and J. J. Pulikkotil$^{1,2}$ }
\affiliation{Academy of Scientific \& Innovative Research (AcSIR), CSIR-NPL, New
Delhi 110012\\
National Physical Laboratory, Council of Scientific \& Industrial
Research, New Delhi 110012}
\begin{abstract}
For iridates with large spatially extended $5d$ orbitals, it may
be anticipated that distant neighbor interactions would play a crucial
role in their ground state properties. From this perspective, we investigate
the magnetic structure of Sr$_{2}$IrO$_{4}$ by including interactions
beyond first and second neighbors, via supercell modeling. Adopting
to first-principles scalar relativistic methods, it is found that
the minimum in total energy among various magnetic structures correspond
to a $\uparrow$$\uparrow$$\downarrow$$\downarrow$ type antiferromagnetic
ordering of the Ir ions for which the magnitude of the electronic
gap, that of the Ir local moments and, the facsimile of the two-peaked
structure in the optical conductivity spectra of Sr$_{2}$IrO$_{4}$
were found to be in good agreement with the experiments. The results
unequivocally evidence that the origin of the electronic gap in Sr$_{2}$IrO$_{4}$
is due to an unconventional antiferromagnetic ordering of Ir ions,
thereby classifying the system as a Slater magnet, rather than the
spin-orbit coupling driven $J_{eff}$ $=$ $\frac{1}{2}$ Mott insulator.
\end{abstract}
\maketitle
Sr$_{2}$IrO$_{4}$ is an insulator at all temperatures \cite{Kim2,Kim1,Jin,Rao,Fisher,Cao,Castaneda,Kini,Klein}
and undergoes an antiferromagnetic transition below $240$ K \cite{Cao,Castaneda,Kini,Klein,Chikara,Crawford,Cava,Shimura}.
Assuming that the strength of spin-orbit coupling (SOC) is comparable
with that of crystal field interaction, Coulomb correlation and Hund's
coupling, a new quantum paradigm has been proposed \cite{Kim2}. In
this model, the crystal field split Ir $t_{2g}$ states are further
split by SOC into a four-fold degenerate $J_{eff}$ $=$ $\frac{3}{2}$
quartet and a two-fold $J_{eff}$ $=$ $\frac{1}{2}$ doublet states.
With Ir in its $+4$ formal valence state, the low energy $J_{eff}$
$=$ $\frac{3}{2}$ states are fully filled with two electrons each,
leaving the $J_{eff}$ $=$ $\frac{1}{2}$ doublet singly occupied.
Furthermore, since the bandwidth of the $J_{eff}$$=$$\frac{1}{2}$
doublets are significantly narrow, Coulomb correlation splits the
doublets into an upper and lower Hubbard band, thereby rendering the
system an insulating ground state \cite{Kim2,Kim3}. The model successfully
accounts for both electron localization and insulating state on equal
footing and derive consistent support from resistivity measurements,
photo-emission spectroscopy, optical conductivity, absorption spectroscopy
and model Hamiltonian based calculations \cite{Kim2,Kim1,Chikara,Takayama}. 

However, few observations had also hinted to itinerant characteristics
in Sr$_{2}$IrO$_{4}$ \cite{Kini,Arita,Li,Yamasaki,Piovera}. Scanning
tunneling microscopy finds that the electronic gap emerges in the
close vicinity of the magnetic transition \cite{Li}, whereas optical
conductivity measurement deduce a strong reduction in the optical
gap with increasing temperature \cite{Moon}. Also, a metal-insulator
transition is observed in the ultrafast dynamics of photo-excited
carriers which indicate to a underlying Slater-type mechanism in Sr$_{2}$IrO$_{4}$
\cite{Hsieh}. Magnetic susceptibility and isothermal magnetization
measurements find the effective paramagnetic moment and the saturation
moment as $0.5$ $\mu_{B}$ and $0.14$ $\mu_{B}$, respectively which
is far less than the expected spin-only value of $1$$\mu_{B}$ for
localized spin of $S$ $=$ $\frac{1}{2}$ \cite{Cao,Kini,Ye}. The
reduction in the magnitude of the Ir moment indicates to strong hybridization
between Ir $5d$ and O $2p$ orbitals. In addition, Sr$_{2}$IrO$_{4}$
displays weak ferromagnetism which is attributed to spin canting \cite{Kim1,Crawford,Kim3,Ye}.
It has been addressed in terms of nontrivial exchange interactions
accounting for the strong coupling of orbital magnetization to the
lattice \cite{Jackeli,Liu1}. The weak ferromagnetism although vanishes
with increasing pressure, system retains the insulating ground state
\cite{Haskel}. The effect is attributed to an increased tetragonal
crystal field thereby substantiating the interplay of structural distortions
and SOC, which affects the balance between the isotropic magnetic
coupling and the Dzyaloshinskii- Moriya anisotropic interaction. It
is also highlighted that distorted in-plane bond angle in Sr$_{2}$IrO$_{4}$
can be tuned through magnetic field \cite{Ge} and epitaxial strain
\cite{Serrao}. Besides, the in-plane anisotropic nature and inter-layer
coupling are also seen to play an important role in the low field
magnetic nature of Sr$_{2}$IrO$_{4}$ \cite{Gim}. Therefore, in
the view of these experimental findings, it is clear that there is
a subtle interplay of SOC, crystal field, Coulomb correlations, magnetic
exchange interactions, and the local chemistry of the underlying IrO$_{6}$
motifs in Sr$_{2}$IrO$_{4}$. 

Significantly, what appears less emphasized in Sr$_{2}$IrO$_{4}$
is the effect of distant near neighbor interactions on the magnetism
and its electronic structure properties. The magnetic structure as
refined from neutron diffraction associates a non-collinear Neel type
ordering of the Ir spins in the crystallographic $a-b$ plane, with
the spin orientation rigidly tracking the staggered rotation of the
IrO$_{6}$ along the $c$-axis \cite{Ye}. However, the Ir $5d$ orbitals
being much extended in space and that they strongly hybridize with
the near neighboring O $2p$ orbitals, it may be anticipated that
the magnetic interactions in the $a-b$ plane would significantly
extend over distant neighbors than those along the $c-$axis. The
antiferromagnetic ordering temperature as high as $240$ K, can be
well thought of one such consequence of distant neighbor magnetic
exchange interactions. 

Here, we present a comprehensive investigation of the electronic and
magnetic structure of Sr$_{2}$IrO$_{4}$, by means of first principles
density functional theory. To include interactions beyond first nearest
neighbors, we model few antiferromagnetic structures on an underlying
super-cell of dimension $2a$$\times$$2a$$\times$$c$, where $a$
and $c$ are the tetragonal lattice parameters of Sr$_{2}$IrO$_{4}$.
Consistent with the previous works, we find that the local approximations
to the exchange correlation potential, such as local density approximation
(LDA) \cite{LDA-PW} and generalized gradient approximation (GGA-PBE)
\cite{GGA-PBE} fail to capture the antiferromagnetic insulating ground
state of Sr$_{2}$IrO$_{4}$. However, using the modified Becke- Johnson
potential (mBJ) \cite{Tran}, we find that the equilibrium corresponds
to an unconventional $\uparrow$$\uparrow$$\downarrow$$\downarrow$
type antiferromagnetic ordering of the Ir ions in the $a$-$b$ plane.
The predictive powers of the calculation are substantiated by the
consistency it yields with the experiments. The magnitude of the insulating
gap and that of the Ir local moment and, the double peak structure
in the materials optical absorption spectra are found to be in good
agreement with the experiments. These findings suggest that the underlying
mechanism that drives Sr$_{2}$IrO$_{4}$ as an antiferromagnetic
insulator is Slater-type, which is in stark contrast with the widely
discussed SOC driven $J_{eff}$ $=$ $\frac{1}{2}$ Mott model. 

Calculations are based on the full potential linearized augmented
plane-wave (FP-LAPW) method as implemented in the Wien2k code \cite{Blaha}.
The lattice parameters were adopted to the experimental values, with
$a$$=5.48$ \AA , and $c=25.83$ \AA{} \cite{Crawford}, and the
position coordinates of the Sr and O ions were fully relaxed. The
ground state properties were obtained using well-converged basis sets
using the Wien2k parameters; $R_{MT}$$K_{max}$$=$ $7$, $G_{max}$$=$$24$
a.u.$^{-1}$and $l_{max}$ $=$$7$ \cite{Blaha}. Additional local
orbitals were also used to account for the semi-core Ir $5p$ states.
The exchange correlation potential to the crystal Hamiltonian was
considered in mBJ formalism \cite{Tran}. 

\begin{figure}
\includegraphics[scale=0.2]{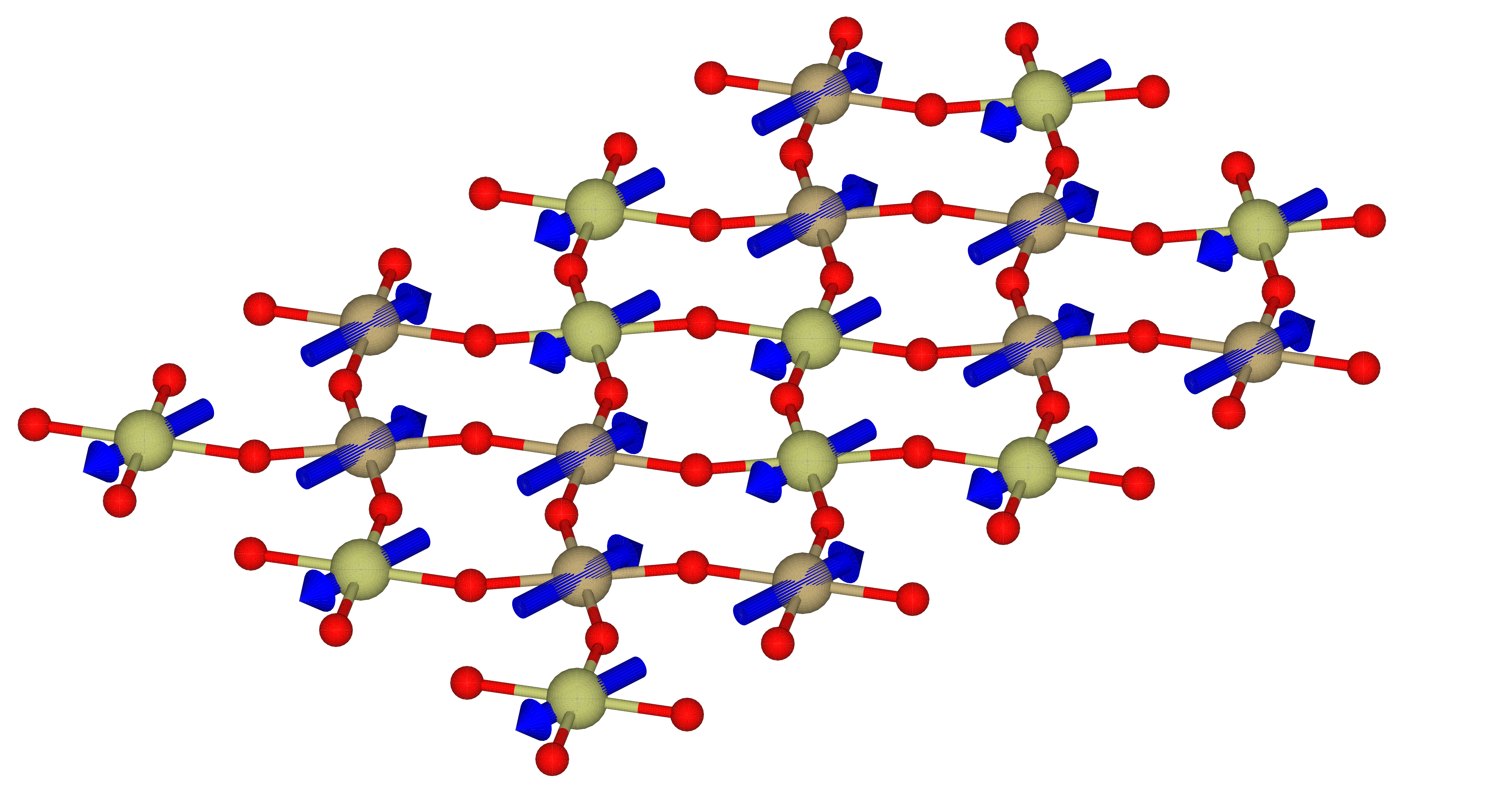}

\caption{\label{FIG_AF3_STRUCTURE}The schematic representation of the AF3
structure showing the antiferromagnetic ordering of Ir moments in
the $a-b$ plane of Sr$_{2}$IrO$_{4}$.}
\end{figure}

Few collinear magnetic structures with different initial Ir spin alignment
were considered in the study. These are shown in Table \ref{TAB_STRUCTURE},
which are described in terms of the Ir spin alignment in the first,
second, third and fourth near neighbors designated as $d_{NN}^{(i)}$;
$i$ $=$ $1$, $4$. Neglecting non-collinearity, AF1 then represents
the experimentally determined structure and FM represents ferromagnetic
ordering. In LDA spin polarized calculations, all structures converged
to a paramagnetic metallic solution. However, in GGA the AF3 spin
configuration converged to an antiferromagnetic metallic solution
with an Ir moment of $0.2$ $\mu_{B}$, while all other structures
converged to a nonmagnetic solution. The AF3 structure was $-1.4$
meV/f.u lower in energy in comparison to its non-magnetic counterpart.
A schematic representation of the AF3 structure is shown in Fig.\ref{FIG_AF3_STRUCTURE}.
The AF3 unit cell consists of $16$ formula units, with an underlying
\emph{Pnna} symmetry of crystal lattice dimensions $a$ $=$ $5.48$$\textrm{\AA}$,
$b=$ $25.83$$\textrm{\AA}$ and $c$ $=$ $10.96$$\textrm{\AA}$.

\begin{table}[h]
\begin{tabular}{>{\raggedright}p{1.2cm}>{\raggedright}p{1.45cm}>{\raggedright}p{1.3cm}>{\raggedright}p{1.3cm}>{\raggedright}p{1.3cm}>{\raggedright}p{1.25cm}}
\hline 
 & Space group & $d_{NN}^{(1)}$ 

($3.88$$\textrm{\AA}$) & $d_{NN}^{(2)}$ 

($5.48$$\textrm{\AA}$) & $d_{NN}^{(3)}$ 

($7.01$ $\textrm{\AA}$) & $d_{NN}^{(4)}$

($7.75$ $\textrm{\AA}$)\tabularnewline
\hline 
\hline 
AF1 & $I$-$4_{2}d$

~ & $4$($\downarrow$) & $4$($\uparrow$) & $4$($\uparrow$) 

$4$($\downarrow$)

~ & $4$($\uparrow$)\tabularnewline
AF2 & $P4_{1}2_{1}2$

~ & $2$($\uparrow$) 

$2$($\downarrow$) & 4($\downarrow$) & $4$($\uparrow$) 

$4$($\downarrow)$

~ & $4$($\uparrow$)\tabularnewline
AF3 & $Pnna$

~ & $2$($\uparrow$) 

$2$($\downarrow$) & $2$($\uparrow$) 

$2$($\downarrow$) & $4$($\uparrow$) 

$4$($\downarrow$)

~ & 4($\downarrow$)\tabularnewline
FM & $I4_{1}/acd$ & $4$($\uparrow$) & $4$($\uparrow$)  & $8$($\uparrow$) 

~ & $4$($\uparrow$)\tabularnewline
\hline 
\end{tabular}

\caption{\label{TAB_STRUCTURE}: Table showing the spin ordering and the space
group of the magnetic structures, generated using $2\times2\times1$
super-cell framework. Here, $d_{NN}^{(i)}$ represents the $i^{th}$
near neighboring shell with respect to a central Ir spin ($\uparrow$)
 ion. The first four near neighboring distances are also shown. }
 
\end{table}

It is well known that the electron density representation of the Coulomb
potential in both LDA and GGA leads to an unphysical self interaction.
As a result, these approximations tend to reduce the self-repulsion
of electrons thereby stabilizing artificially delocalized electronic
states \cite{Cohen,Mori}. Among various correction schemes that have
been proposed \cite{Heyd,Bechstedt,Georges}, we adopt to the mBJ
formalism. With $t$ and $\rho$ representing the kinetic energy density
and electron density, respectively, a screening term of the form $\sqrt{\frac{t}{\rho}}$
is introduced in the mBJ exchange potential, the contribution of which
is calculated by $\frac{\left|\nabla\rho\right|}{\rho}$ \cite{Tran}.
As a result, regions with low density are associated with higher positive
potential thereby increasing the energy of these states \cite{Tran,Koller}.
The mBJ formalism is applicable for Sr$_{2}$IrO$_{4}$ and also for
other iridates \cite{Vijeta2,Vijeta1} since the states in the vicinity
of Fermi energy are predominantly anti-bonding in character. It should
be noted that the anti-bonding orbitals have less electron density,
thus the choice of mBJ exchange potential for iridates is well justified. 

\begin{figure}
\includegraphics[scale=0.33]{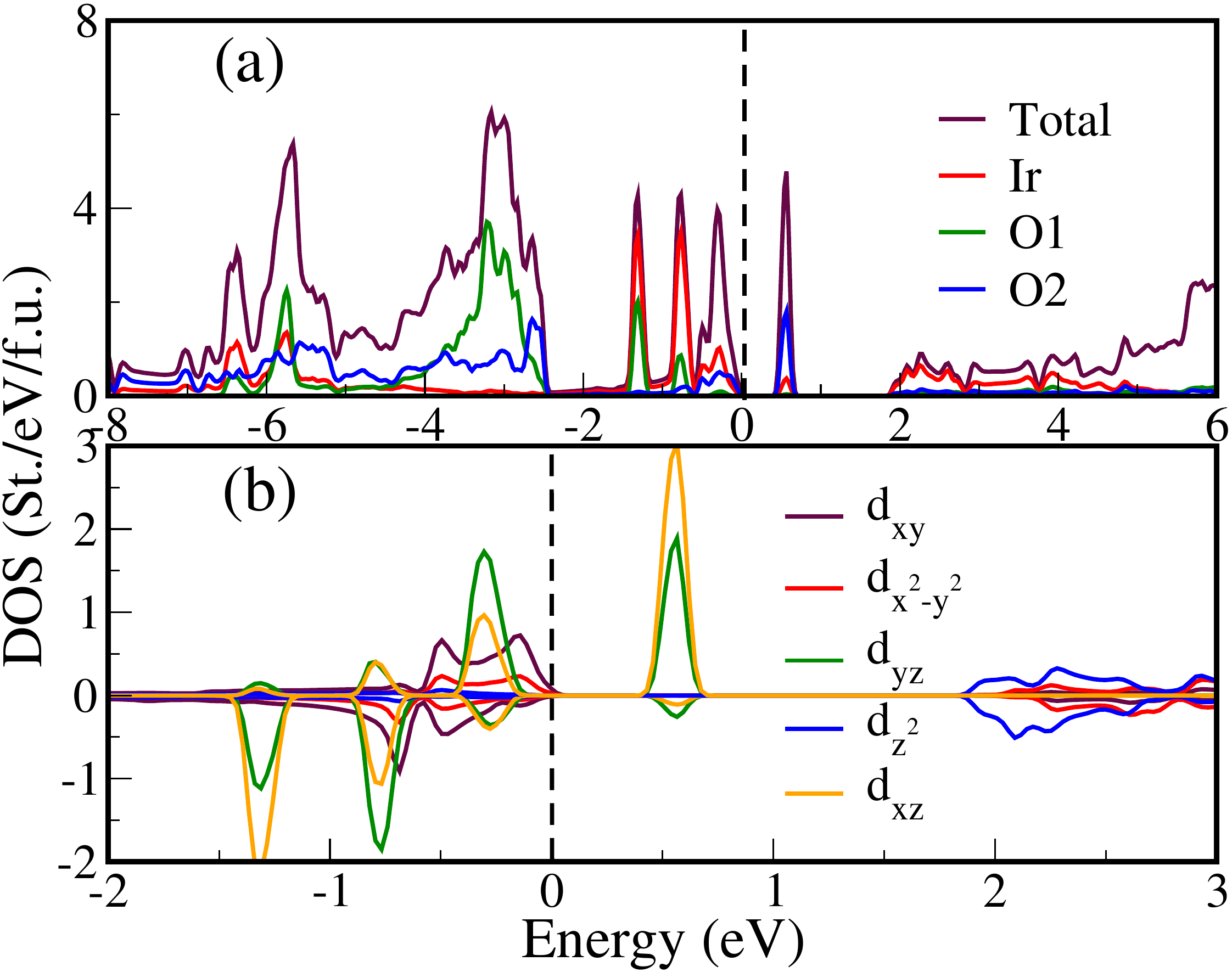}

\caption{\label{DOS-AFM-MBJ}(color online): The mBJ density of states of the
AF3 antiferromagnetic structure. (a) Total and atom resolved partial
density of states for the AFM unit-cell and, (b) Ir $5d$ resolved
partial density of states per Ir atom. Here, the O1 and O2 atoms represent
the apical and in-plane O atoms, respectively.\textbf{ }The broken
line through energy zero represents the reference Fermi energy. }
\end{figure}

In Fig.\ref{DOS-AFM-MBJ}, we show the mBJ generated total, atom resolved
and Ir $5d$ resolved density of states (DOS) of Sr$_{2}$IrO$_{4}$
with AF3 spin configuration in the Ir sub-lattice. The spectra reveal
Sr$_{2}$IrO$_{4}$ to be an insulator with an electronic gap of $0.47$
eV, consistent with the experimental value of $0.54$ eV \cite{Moon}.
Here, we note that the magnitude of the insulating gap in Sr$_{2}$IrO$_{4}$
have been reported between $0.1$$-$$0.6$ eV, with the lowest determined
from the resistivity data fit using a thermal activation model \cite{Shimura,Ge}
and also from the earlier GGA+U+SOC calculations \cite{Kim2,Jin}.
The highest value of the insulating gap of $0.6$ eV follows from
the density of states measurements using the scanning tunneling spectroscopy
\cite{Dai}. Intermediate values of the gap are reported from the
optical conductivity, resonant inelastic x-ray scattering \cite{Kim2,Jin,Kim3}
and also from the photoemission spectroscopy measurements \cite{Kim2,Wang,Wojek}. 

In accordance with the ionic model which associates Ir $5d$ manifold
with five electrons, the electronic gap is found to reside well within
the Ir $t_{2g}$ manifold which extends over the range $E$(eV) $\subset$
$\left[-1.5,\,0.8\right]$. Four distinct localized features, which
are predominantly of Ir $d_{xz}$ / $d_{yz}$ orbital characters are
observed in the spectra. Three of them are in the occupied part of
the spectra centered at $-1.32$ , $-0.77$ and $-0.31$ eV below
$E_{F}$ and, the fourth peak at $0.55$ eV above $E_{F}$. The position
of these bands indicate to three Ir $t_{2g}$ inter-band transitions
with the first, second and third transition energies being $\simeq$
$0.86$ eV , $1.32$ eV and $1.87$eV, respectively. These energies
are reasonably in good agreement with the optical conductivity measurements,
where two transition peaks labeled $\alpha$ and $\beta$ were determined
to be at $\simeq$ $0.5$ eV and $1$ eV, respectively \cite{Moon,Sohn}. 

However, unlike the $d_{xz}$ / $d_{yz}$ states, the $d_{xy}$ states
appear relatively more widespread on the energy scale. We note that
the Ir-O distance in Sr$_{2}$IrO$_{4}$ corresponds to $1.98$ $\textrm{ \AA}$
and $2.04$ $\textrm{\AA},$ for in-plane and apical O ions, respectively.
For the in-plane O2 ions, the $2p_{z}$ orbitals hybridize with the
Ir $d_{xz}$ and $d_{yz}$ orbitals, while the O $2p_{x}$ and $2p_{y}$
mix with the $d_{xy}$, $d_{z^{2}}$ and $d_{x^{2}-y^{2}}$ orbitals.
On the other hand, for apical O1 ions the $2p_{z}$ orbitals hybridize
with the Ir $d_{z^{2}}$ and the $2p_{x}$ / $2p_{y}$ mix with the
$d_{xz}$ / $d_{yz}$ states. Thus, the crystal chemistry suggests
a mixing of the Ir $t_{2g}$ and $e_{g}$ states in Sr$_{2}$IrO$_{4}$
primarily due to the rotation of the IrO$_{6}$ octahedra. The rotation
of the octahedra mixes the otherwise orthogonal Ir $d_{xy}$ and $d_{x^{2}-y^{2}}$
orbitals and consequently push the $d_{xy}$ states below the Fermi
energy. The $d_{xy}$ and $d_{x^{2}-y^{2}}$ hybridization also results
in a pseudo-gap like feature which is manifested $\simeq$$-0.6$
eV below $E_{F}$. Further, the valence band energy integration of
the orbitals states showed that the $d_{xy}$+ $d_{x^{2}-y^{2}}$
orbitals are occupied with $\simeq$ $2$ electrons per Ir ion, while
the electron occupation in the $d_{xz}$+$d_{yz}$+$d_{z^{2}}$ sums
to $3$ electrons per Ir ion, with $d_{xz}$ and $d_{yz}$ occupancy
being $1.15$ and $1.29$ electrons, respectively. Also, the integrated
DOS of the $d_{xy}$ / $d_{yz}$ orbitals above $E_{F}$ was determined
to be $\simeq$$1$ e$^{-}$per Ir ion. Thus, we find that the scalar
relativistic calculations with exchange potential as described in
the mBJ formalism predicts Sr$_{2}$IrO$_{4}$ to be an antiferromagnetic
insulator. 

The magnitude of the local magnetic moment at the Ir sites in the
AF3 structure was calculated as $\simeq$ $0.57$ $\mu_{B}$. The
value is significantly higher than those determined from experiment,
the latter which deduce the value as $0.2$ $\mu_{B}$ \cite{Ye}.
The overestimation of the Ir local moment might be due to the PBE-GGA
functional in the calculation. However, the Ir magnetic moment is
found to be much smaller than spin only value of $1$ $\mu_{B}$ anticipated
for a $S$ $=$ $\frac{1}{2}$ system. This may be partly attributed
to the strong hybridization of the Ir $5d$ $-$ O $2p$ orbitals.
The effects of hybridization are also manifested on the induced moments
at the O sites. We note that the AF3 structure has a $\uparrow$$\uparrow$$\downarrow$$\downarrow$
type antiferromagnetic ordering of Ir ions in the $a-b$ plane of
the tetragonal unit cell. For those in-plane O ions which bridge the
Ir ions in the $a$-$b$ plane with same polarization, \emph{i.e.,
}$\uparrow$$\uparrow$ or $\downarrow$$\downarrow$ the induced
magnetic moment is calculated as $\simeq$ $0.12$ $\mu_{B}$ while
for oppositely polarized Ir ions the moment is $\simeq$ $0.06$ $\mu_{B}$
. The induced moments on the apical O ions were found to be $0.03$
$\mu_{B}$. 

One of the well accepted methods to validate the electronic structure
is by its optical response. In experiments, a double-peak structure
with maxima around $0.5$ eV and $1$ eV have been observed, with
the former peak being relatively sharper than the latter \cite{Kim3,Sohn,Lee}.
These peaks are associated with two Ir $d-d$ transitions, which in
terms of the $J_{eff}$ model are due to the transitions from occupied
$J_{eff}$ $=$ $\frac{3}{2}$ and $\frac{1}{2}$ states to the unoccupied
$J_{eff}$ $=$ $\frac{1}{2}$ states. The spectra has been well reproduced
by the LDA+U+SO calculations, thereby suggesting the importance and
interplay of SOC and Coulomb correlations in Sr$_{2}$IrO$_{4}$ \cite{Kim2}.

\begin{figure}
\includegraphics[scale=0.3]{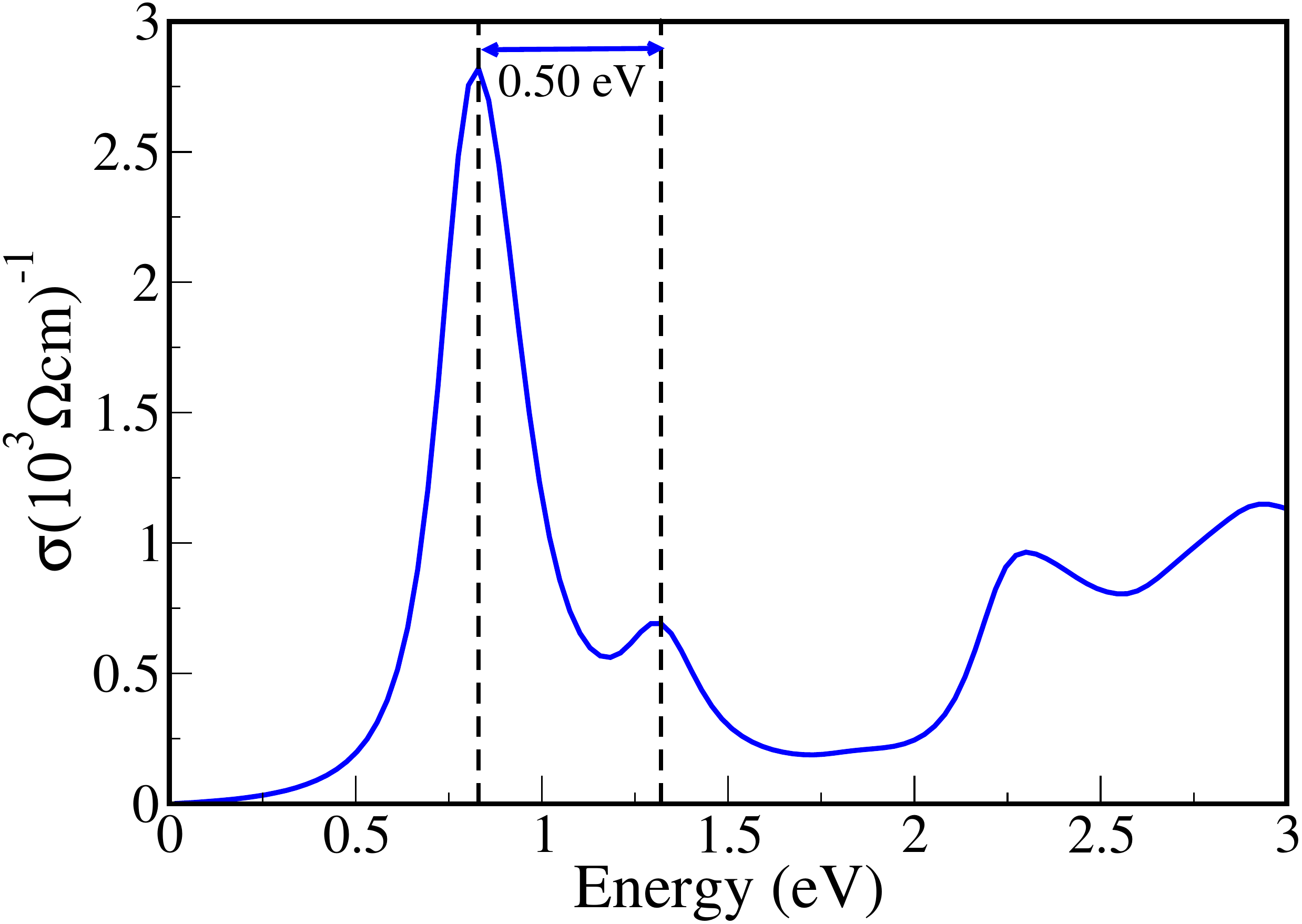}

\caption{\label{Optical-conductivity}The calculated optical conductivity spectra
of Sr$_{2}$IrO$_{4}$ using the scalar relativistic Hamiltonian with
an underlying AF3 structure. Two peaks correspond to Ir $d-d$ transitions
and are positioned at $0.82$ eV and $1.32$ eV, respectively. The
energy difference between the peaks corresponds to a magnitude of
$0.5$ eV.}
 
\end{figure}
In Fig. \ref{Optical-conductivity}, we show the optical conductivity
calculated for Sr$_{2}$IrO$_{4}$ with the underlying AF3 structure.
Consistent with the experimental spectra, we obtain two characteristic
peaks, centered on the energy scale\emph{ }at $\simeq$ $0.82$ eV
and $1.32$ eV, respectively. We note that the position of the peaks
are shifted to higher energies in comparison with experiments \cite{Sohn,Propper}
which is primarily due to the larger band gap estimated ($0.57$ eV)
in our calculations. However, what is very consistent between the
experiments and that of our results is the energy difference between
the two peaks, which is found to be $0.5$ eV. Our results, therefore
show that the origin of electronic gap in Sr$_{2}$IrO$_{4}$ is associated
with the long ranged antiferromagnetic interactions, and hence a Slater-insulator. 

\begin{figure}
\includegraphics[scale=0.33]{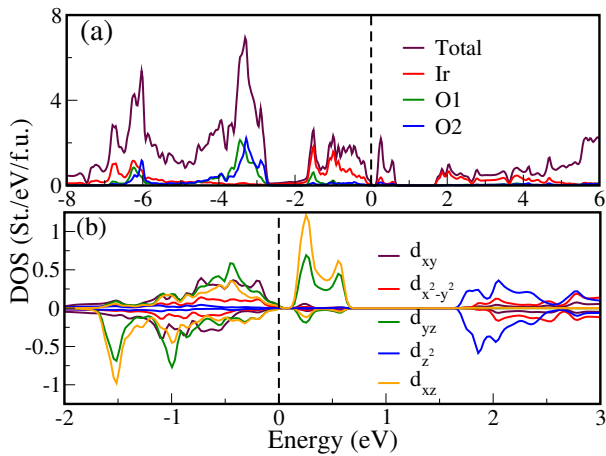}

\caption{\label{DOS-AFM-MBJ-SOC}(color online): The mBJ-GGA+SOC density of
states of the AF3 antiferromagnetic structure. (a) Total and atom
resolved partial density of states for the AFM unit-cell and, (b)
The Ir $5d$ resolved partial density of states. Here, the O1 and
O2 atoms represent the apical and in-plane O atoms, respectively.\textbf{
}The broken line through energy zero represents the reference Fermi
energy. }
\end{figure}
We also investigated the effect of spin-orbit coupling (SOC) on the
electronic structure of Sr$_{2}$IrO$_{4}$. The GGA-mBJ+SOC density
of states are shown in Fig.\ref{DOS-AFM-MBJ-SOC}. In the calculation
the SOC was included for the valence states through the second variational
step with the mBJ scalar relativistic basis, where states up to $10$
Ry above $E_{F}$ were included in the basis expansion. While the
overall features of bonding states are more or less unaltered, we
find noticeable changes in the anti-bonding region. The Ir $5d$ bands
are more broadened manifesting an enhanced hybridization of the $t_{2g}$
states with the O $2p$ orbitals. The hybridization not only is found
to reduce the band gap to $0.17$ eV, but also decreases the magnitude
of the Ir local magnetic moment to $0.47$ $\mu_{B}$. 

\begin{figure}
\includegraphics[scale=0.33]{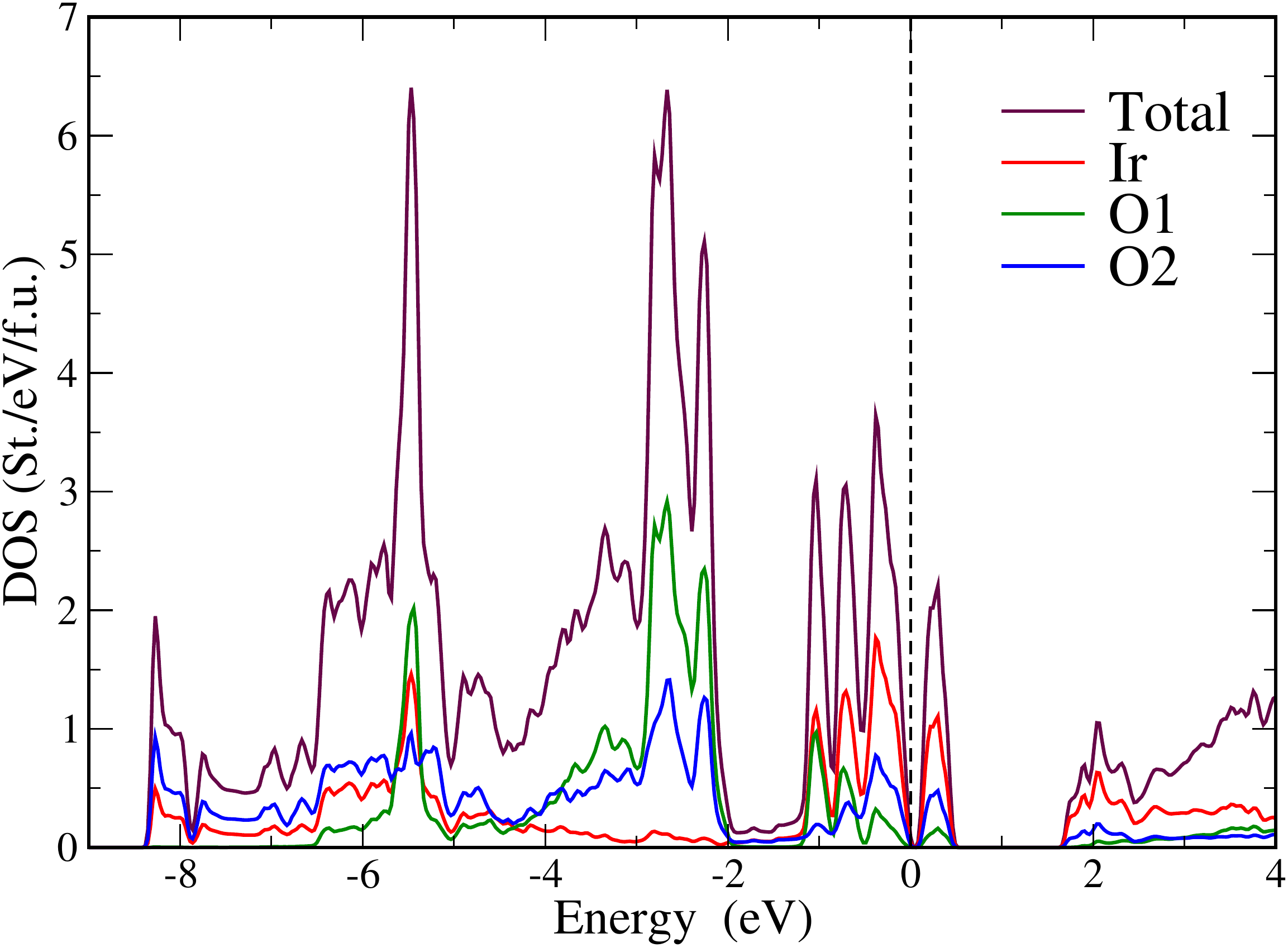}

\caption{\label{DOS-AFM-GGA_U2}(color online): The density of states of the
AF3 antiferromagnetic structure calculated in the GGA+U$_{eff}$ (U$_{eff}$
$=$ $2$ eV) Hamiltonian, showing the total and atom resolved partial
density of states. Here, the O1 and O2 atoms represent the apical
and in-plane O atoms, respectively.\textbf{ }The broken line through
energy zero represents the reference Fermi energy. }
\end{figure}
So to check whether the insulating gap is originally due to the unconventional
antiferromagnetic ordering of Ir spins and not pertained to the choice
of the exchange functional, we also performed GGA+U$_{eff}$ calculations,
with $U_{eff}$ $=$ $2$ eV. Quite interestingly, the overall features
in the density of states (Fig.\ref{DOS-AFM-GGA_U2}) were found very
much similar to that obtained with the mBJ-GGA. However, the calculated
band gap and Ir local moment was $0.11$ eV and $0.41$ $\mu_{B}$,
respectively. In general, our results following a comprehensive set
of calculations concisely and convincingly show that SOC is lesser
significant interaction in rendering Sr$_{2}$IrO$_{4}$ an antiferromagnetic
insulating ground state. 

In summary, using the first-principles density functional theory based
scalar relativistic calculations with exchange potential described
in mBJ formalism, we find that Sr$_{2}$IrO$_{4}$ is an unconventional
Slater-type antiferromagnetic system. The calculated magnitude of
the electronic gap, that of the Ir local moment and, the two peak
structure in the materials optical conductivity are found to be very
consistent with the experimental results. Contrary to the present
understanding that Sr$_{2}$IrO$_{4}$ is a SOC driven $J_{eff}$
Mott insulator, our calculations show that the role of of SOC in Sr$_{2}$IrO$_{4}$
is of lesser significance in rendering the system its insulating ground
state. Our results, which are based on density functional theory,
are expected to stimulate further experimental works with an objective
to unravel the magnetic structure of the system and the nature of
Ir magnetism, thereby providing robust understanding of iridates,
in general.

\end{document}